\documentclass[prb,twocolumn,amssymb,amsmath,aps,floatfix,makecell]{revtex4}
\usepackage[colorlinks=true,linkcolor=blue,citecolor=blue,urlcolor=blue]{hyperref}
\usepackage{graphicx,bm}
\usepackage{epstopdf}
\usepackage{xcolor}

\begin{document}

\title{Hall effect in Poiseuille flow of two-dimensional electron fluid }

\date{\today}

\author{A.~N.~Afanasiev}
\author{P.~S.~Alekseev}
\author{A.~A.~Danilenko}
\author{A.~P.~Dmitriev}
\author{A.~A.~Greshnov}
\author{M.~A.~Semina}
\affiliation{Ioffe Institute, St.Petersburg 194021, Russia}

\begin{abstract}

The hydrodynamic regime of charge transport  has been recently realized in high-quality conductors. In the hydrodynamic as well as in the Ohmic  regimes the main part of the Hall resistance of a long sample is  determined by the balance between the Lorentz force and the electric force, acting on conduction electrons. Experimentally observed deviations of the Hall resistance in hydrodynamic samples  from such the ``standard'' value are usually associated with the Hall viscosity term in the Navier-Stokes equation. In this work we theoretically study the Hall effect in a Poiseuille flow of a two-dimensional electron fluid. We show that the near-edge semiballistic layers with the  width of the order of the inter-particle mean free path, which inevitably appear near sample edges, give the contribution to the Hall resistance which is comparable with the bulk contribution from the Hall viscosity. In this way, the measured deviations of the Hall resistance from the ``standard'' one in hydrodynamic samples by  the usual contact techniques should be associated with both the Hall viscosity in the bulk and the semiballistic effects  in the near-edge layers.

\end{abstract}

\maketitle


\section{ Introduction}
Hydrodynamic  collective behavior of electron systems, in which  interparticle collisions are  the dominant scattering mechanism, has been studied theoretically for a very long time~\cite{Gurzhi1968,Hruska2002}. However, the experimental feasibility of the hydrodynamic electron transport regime has became possible only in the last few years with the emergence of high-quality samples of graphene~\cite{Bandurin2016,Levitov2016, Kumar2017, Berdyugin2019, Sulpizio2019,Ku2020,Polini2020}, quasi-two dimensional metals~\cite{Moll2016}, Weyl semimetals~\cite{Gooth2018}, and GaAs quantum wells~\cite{Dai2010,Hatke2011,Bockhorn2011,Hatke2012,Mani2013,Shi2014,Bialek2015,Gusev2018_AIP,Levin2018,Gusev2018_PRB}. The formation of a viscous  flow of the electron fluid in these experiments manifests itself  in the emergence of the negative nonlocal resistance~\cite{Bandurin2016,Levitov2016,Gusev2018_PRB}, the giant negative magnetoresistance~\cite{Dai2010,Bockhorn2011,Hatke2012,Mani2013,Shi2014,Alekseev2016,Gusev2018_AIP,Levin2018,Gooth2018}, the magnetic resonance~\cite{Dai2010,Hatke2011,Bialek2015,Alekseev_Alekseeva_2019} at the double cyclotron frequency   $\omega=2\omega_c$,  and the specific dependence of the mean sample resistivity on its width~\cite{Moll2016}. The recent breakthrough in the experimental techniques allowed to study the very process of formation of the hydrodynamic regime from the ballistic and the Ohmic ones. Namely, in Refs.~\cite{Sulpizio2019,Ku2020} direct measurements of the evolution of the Hall electric field and the current density profiles in a graphene stripe allowed to trace the ballistic-hydrodynamic and Ohmic-hydrodynamic transitions with the variation of temperature and magnetic field. Those effects were theoretically explained and described  in~Refs.~\cite{Guo2017,Scaffidi2017,Holder2019,Afanasiev2021,Afanasiev2021_2}.

Studies of the Hall effect in bulk conductors often provides important information about the type of transport and the microscopic nature of charge carriers. In the case of the hydrodynamic transport, a non-trivial size-dependent contribution in the Hall resistance, additional  to the ``standard'' one, $ \rho^{0}_{xy}=B/(n_0ec)$, arises due to the non-diagonal Hall viscosity term in the Navier-Stokes equation. This contribution  is considered to be an important fingerprint of the formation of the electron fluid. Its measurements were reported for graphene samples~\cite{Berdyugin2019} and high-mobility GaAs quantum wells~\cite{Gusev2018_PRB}.  Numerical theory~\cite{Scaffidi2017} yields a small negative correction to $\rho^{0}_{xy}$ in long samples with rough edges. This prediction agrees with the experimental observation~\cite{Gusev2018_PRB}  by its sign. However, we consider that more detailed theoretical studies
are necessary
in order to clarify whether
the size-dependent correction to $ \rho^{0}_{xy}$
observed in Refs.~\cite{Berdyugin2019,Gusev2018_PRB} is  indeed originates  only due to  the Hall viscosity.
For example,  it was mentioned in Ref.~\cite{Alekseev2016} that such  Hall viscosity correction  $\Delta \rho^{h}_{xy}  $
in a conventional Poiseuille flow
 is estimated as the squared parameter of applicability of hydrodynamics,  $l_2/W \ll 1$, times  $\rho^{0}_{xy}$, where $l_2$ is the relaxation length   of the viscous shear stress  and $ W$ is the sample width.   In this connection, a non-trivial question arises what is the ratio of  the Hall viscosity contribution, $\Delta \rho^{h}_{xy} \sim (l_2/W )^2\rho^{0}_{xy}$, in  the Hall resistance $\rho_{xy}$ and the contribution in  $\rho_{xy}$ from the near-edge layers of  the widths $\sim l_2 $. In such layers
 the flow is partially formed by the electrons reflected from edges and, thus, is partially ballistic. The purpose of the current work is to address this question.

The near-edge semiballistic layers in a flow of two-dimensional electrons have been recently theoretically studied in Refs.~\cite{Kiselev2019,Raichev2022} for rough and for slightly curved sample edges.  It was shown that an approximate solution of the kinetic equation in such layers leads to refined boundary conditions for the mean electron velocity  at the sample edges. They  allow to solve the hydrodynamic Navier-Stokes equations in the main and the first orders by  the small parameter $l_2/W$. Those  conditions contain the slipping length, $l_{sl}\sim l_2 $, that effectively accounts the semiballistic flow in the near-edge regions and lead to the first order corrections by $l_2/W$ to the velocity profile.

In this work we study the influence of the near-edge semiballistic layers on the Hall effect in a Poiseuille flow of two-dimensional electrons. We consider the case of small magnetic fields in which  the cyclotron radius $R_c$ is much larger than the relaxation length  $l_2$.  We construct the two different models of the semiballistic flow in the near-edge layers.

The first model is based on a simplified picture  of ``virtual'' sharp interfaces between the  near-edge semiballistic layers and the bulk purely hydrodynamic region of the flow (see Fig.~1). An approximate analytical solution of the kinetic equation in the sharply separated layers, analogous to the solution constructed in  Ref.~\cite{Kiselev2019}, is derived.  Such solution allows to obtain non-trivial boundary conditions for the Navier-Stokes equation, which governs the flow in the bulk region. The semiballistic solution  also yields the refined profiles of the current density and the  Hall electric field in the near-edge layers.

The second model is based on the solution of the kinetic equation in the whole sample for a truncated distribution function which contains the hydrodynamic part and the part describing ballistic effects. The last one is roughly described  by the  third-order  angular harmonics in the Fourier decomposition of the distribution function by the electron velocities. This function effectively describes the semiballistic flow in the near-edge layers, therefore it  leads to the non-trivial  boundary conditions for the Navier-Stokes equation and to the refined    expressions for the mean velocity and
 the Hall electric field.

Both these two models yield corrections  $\Delta \rho_{xy}$ to the Hall resistance  relative to its ``standard'' value  $\rho^{0}_{xy}$.
 The magnitude of $\Delta \rho_{xy}$ turns out of the order of the Hall viscosity correction, $\Delta \rho^{h}_{xy} \sim (l_2/W )^2\rho^{0}_{xy}$.
Based on this result, we conclude that experimental studies of the Hall effect in hydrodynamic samples by the usual measurements of the total current and the voltages between contacts  should give information
not about the very Hall viscosity, but about  the properties of both the bulk hydrodynamic region, where the flow is affected by the Hall viscosity,
and the near-edge semiballistic  layers. The relative importance of these two regions may depend on geometry of the system.


\section{  Conventional hydrodynamics of Poiseuille flow of viscous electron fluid }
\begin{figure}[t!]
	\includegraphics[width=0.8\linewidth]{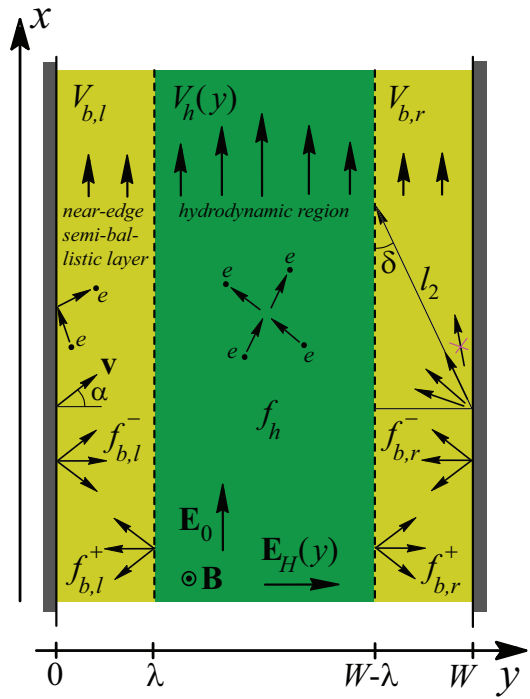}
	\caption{\label{Fig1} The proposed model ``sharp-edge'' of the Poiseuille flow in a wide sample, $W \gg l_2 $, with rough edges. The flow is divided on the bulk hydrodynamic region  $\lambda <y< W- \lambda $  (here $\lambda \sim l_2 $)  and   the ballistic near-edge layers $0<y<\lambda$, $W- \lambda <y<W$ with sharp interfaces at $y=\lambda,\,W-\lambda$. In the first region only the electron-electron scattering takes place, while in the second layers both  electron-electron collisions and collisions with the edges occur. For simplicity we consider  that the electron-electron scattering in the near-edge layers for electrons with the velocity angles being close to  the $x$ axis, $ |\pi/2 - |\varphi||<\delta (\lambda) $  is infinitely fast [here $ \sin \delta = \lambda /l_2 $]. }
\end{figure}

We consider a flow of two-dimensional electrons in a long sample with rough edges in an external electric field $\mathbf{E}_0$ directed along the sample (see Fig.~\ref{Fig1}). A magnetic field $\mathbf{B}$ is applied perpendicularly to the layer of two-dimensional electrons. It is considered to be  classically weak: $\omega_c\tau_2\ll 1$. Here $\omega_ c$ is the cyclotron frequency of two-dimensional electrons and $\tau_2$  is the relaxation time of the second angular moment of the distribution function, which determines the shear stress in the fluid flow.

If the sample width $W$ is much larger than the shear stress relaxation length $l_2 = v_D \tau_2$, a macroscopic description for the flow in the sample can be applied ($v_F$ is the Fermi velocity). The average velocity of the flow is then determined by the Navier-Stokes equations with the magnetic-field dependent diagonal viscosity $\eta_{xx}$ coefficient and no-slip boundary conditions~\cite{Alekseev_Semina_2018,Alekseev_Semina_2019}. For the geometry of the Poiseuille flow, assuming the velocity is directed along the $x$ axis and depends on the $y$ coordinate along the sample section, we have:
\begin{equation}
\label{NS_eq}
\left\{
\begin{array}{l}
   \eta_{xx} V_{h0}'' + e E_0  / m = 0  \:,
\\\\
   \omega_c V_{h0} + \eta_{xy} V_{h0}''  + e E_H (y) / m = 0
\:,
\end{array}
\right.
\end{equation}
where the symbol ``$\,'\,$'' denotes the derivative by $y$, $E_H(y)$ is the Hall electric field, and
\begin{equation}
 \label{etas}
\left.
	\begin{array}{c}
   \eta_{xx}
   \\
   \eta_{xy}
   \end{array}
   \right\}
=
 \frac{ v_F^2 \tau_2 / 4 }{1+4\omega_c^2 \tau_2^2}
\left\{
	\begin{array}{c}
  1
\\
  2\omega_c \tau_2
\end{array}
\right.
\end{equation}
are  the diagonal and the non-diagonal (``Hall'') viscosities in the presence of magnetic field.  We use the subscript ``$h$'' for velocity and other values in order to, first,  denote the values calculated within the purely hydrodynamic description of the current section, and, second, to  distinguish the values  in the hydrodynamic region of the flow from the values in the near-edge semiballistic regions considered below (see Fig.~\ref{Fig1}).

The diffusive scattering of electrons on rough edges in the simplest purely hydrodynamical  approach is described  by the ``no-slip'' boundary conditions:
\begin{equation}
 \label{bond_c_simpest}
    V_{h0}|_{ y=0, \, W } = 0
   \:.
\end{equation}
These conditions describe the absolute sticking of the flow to the boundaries.

The solution of equations (\ref{NS_eq}) with conditions (\ref{bond_c_simpest}) yields the description of the flow  in wide samples, $W \gg l_2 $, in the main order by the ratio $l_2/W$. One easily obtains a parabolic velocity profile of the two-dimensional Poiseuille flow:
\begin{equation}
 \label{V_H_simplest}
    V_{h0}(y) = \frac{eE_0}{2 m \eta_{xx}} \, y \, (W-y)
   \:,
\end{equation}
leading to the electric  current:
\begin{equation}
\label{II}
     I=  \frac{ e^2 n_0 E_0 W^3 }{12 \, m \, \eta_{xx} }  \:.
\end{equation}
The Hall electric field takes the form:
\begin{equation}
\label{E_H_hydr}
    E_H^h(y) = - \frac{m  \omega_c }{e} V_{h0}(y) + \frac{ \eta_{xy} }{ \eta_{xx} } \, E_0
 \:
  \:,
\end{equation}
that corresponds to  the Hall voltage $  U_H  ^h = -  \int_0^W dy\, E_H^h(y)  $ between the edges:
\begin{equation}
  \label{U_H}
     U_H  ^h=  \frac{ m \omega_c }{e^2}\, \frac{ I}{n_0} - \frac{ \eta_{xy} }{ \eta_{xx}  } \, E_0 \,  W \: .
\end{equation}

Below in this work we calculate the transport characteristics on the limit $\omega_c \tau_2 \ll 1 $ in the zero and the first orders by $\omega_c $. In this limit
the current and the longitudinal resistances are independent on magnetic field and the Hall field is linear in magnetic field.

The above expressions for the velocity, the current, the Hall field, and the Hall voltage via the values $v_F$ and $\tau_2 $ at  $\omega_c \tau_2 \ll 1 $ are presented in appendix A.
The resulting  Hall resistance $\varrho _{xy} = U_H/I $ consists of the standard kinematic term, $\rho_{xy}^0 = B/(n_0ec)$, similar to case of the Ohmic transport in disordered bulk samples  at zero temperature, and the bulk hydrodynamic correction proportional to the Hall viscosity:
\begin{equation}
   \label{Hall_re_corr_hydr}
   \frac{   \Delta \varrho_{xy}^h   }{ \rho_{xy}^0 }= -  \frac{ 12 \, \eta_{xy} }{ \omega_c W^2 }
   =
    - \frac{ 6 \,  l_2 ^2 }{ W^2 }
   \: ,
\end{equation}
This result was obtained and discussed in Refs.~\cite{Alekseev2016,Scaffidi2017}.


\section{  Semiballistic near-edge layers in  Poiseuille flow }
\label{Sec:Semiballistic}
In this section we develop an approximate description of flow  accounting for the hydrodynamic bulk  region and the semiballistic near-edge layers with virtually sharp boundaries (see Fig.~1).

First, following to Refs.~\cite{Alekseev_Semina_2018,Alekseev_Semina_2019,Afanasiev2021,Yu_Alekseev_AP_Dmitriev} we  will perform the semiballistic transport of two-dimensional electrons in the near-edge layers in a weak magnetic field~$B$. We will derive an estimation for the ballistic part of the distribution function in these layers in the zero and the first orders by a weak magnetic field,  $\omega_c \tau_2  \ll 1 $. From the zero-order part   of such ballistic-hydrodynamic  distribution, we will obtain  the generalized boundary conditions for the average velocity accounting for the flow in both the bulk hydrodynamic and the near-edge ballistic regions (see Fig.~\ref{Fig1}). Then we will solve the Navier-Stokes equations for the velocity with the obtained boundary condition that leads to a correction to the current by the small parameter $l_2/W$. Second, using the first-order by $B \to 0 $ correction to the constructed distribution function, we will estimate the contribution to the Hall voltage from the near-edge semiballistic regions.

The electron distribution function in a weak electric field has the form:
\begin{equation}
  \label{f}
    f_\mathbf{p}(y) \, =  \, f_F(\varepsilon _ p ) \, - \, f'_F(\varepsilon _ p )  \, f(y,\alpha)
    \, ,
\end{equation}
where $f_F$ is the Fermi function, $\varepsilon _ p  = p^2/(2m)$ is the electron energy spectrum,    $\mathbf{p}=p \, (\sin\alpha, \cos\alpha)$  is the electron momentum,
and
 the function $f(y,\alpha)$ is proportional to the electric field. The kinetic equation takes the form:
\begin{equation}
 \label{kin}
	v_F \cos\alpha \,
       \frac{\partial \tilde{f}}{\partial y} - \omega_c \,  \frac{\partial \tilde{f}}{\partial \alpha} - e E_0 v_F \sin\alpha \,  = \, 0
    \: .
\end{equation}
Hereinafter $\tilde{f} $ is the generalized distribution function:
\begin{equation}
 \label{f_tilda}
    \tilde{f} (y,\alpha) \, = \,  f (y,\alpha) \, + \, e \phi(y)
     \:,
\end{equation}
where $\phi$ is the electrostatic potential of the Hall field $E_H=-\phi'$. Equation~(\ref{kin}) neglects the electrons whose last scattering event  was a collision with other electrons. Below we just write $f \equiv \tilde{f} $ for brevity.

In the near-edge regions with the widths of the order of~$l_2$, the last scattering event for a substantial fraction of electrons before leaving these regions is the reflection from the sample edges. In other words, the flow in these layers is semiballistic. It is described by the kinetic equation in which the collision term can be neglected in a rough approximation.

For rough sample edges, the boundary conditions for $f$ state that the electrons incident on the edges  are diffusely scattered by the  edge roughness, so for the electron reflected from the left edge  with the angles $ - \pi/2 < \alpha < \pi/2 $, we have:
\begin{equation}
 \label{diff_b_c}
\begin{array}{c}
\displaystyle
f_{b,l}^{+}(0,\alpha)
\displaystyle
= -\frac{1}{2}\int\limits_{\pi/2}^{3\pi/2} f_{b,l}^{-}(0,\alpha') \cos \alpha' d\alpha'
\: .
\end{array}
\end{equation}
At the right edge the similar condition  for the reflected  electrons with the angles~$\pi/2 <  \alpha < 3\pi/2$~is:
\begin{gather}
\label{diff_b_c2}
	f_{b,r}^{-}(W,\alpha)=\frac{1}{2}\int\limits_{-\pi/2}^{\pi/2} f_{b,r}^{+}(W,\alpha') \cos \alpha' d \alpha' \: .
\end{gather}
Here the notations $ f_{b,l}^{+} $ and $ f_{b,l}^{-}$ denote the semiballistic distribution functions (``$b$'') of the outgoing electrons near the left (``$l$'') edge with $v_y>0 $ (``$+$''), and the incoming ones with $v_y<0$ (``$-$''), respectively. The functions $f_{b,r}^{\pm}$ are defined analogously (in particular, ``$r$'' denotes the right edge). Such boundary conditions are strict for completely rough edges. The total distribution functions  $f_{b,l}$ and $f_{b,r}$ in the semiballistic layers, consisting of the functions $ f_{b,l}^{+} $, $f_{b,l}^{-} $  and $f_{b,r}^{+} $, $f_{b,r}^{-}$, respectively, have discontinuities at $\varphi = \pi/2$ and $\varphi = 3\pi/2$.

Boundary conditions~(\ref{diff_b_c}) and (\ref{diff_b_c2}) lead to the vanishing of ``$y$'' component of the the $y$-component of the mean velocity at the edges:  $\left. V_{y,b} \right|_{y=0, \, W} =0$. Owing to the conservation of the number of electrons, from these boundary conditions we obtain  $V_y (y) =0 $  everywhere in the sample.

In the near-edge regions,
instead of the locally equilibrium hydrodynamic velocity $V_h(y)$, we should use the averaged velocity $ V_{x,b}(y) \equiv V_b(y)$. It is determined by the distribution function $f_{b}$ by the formula:
\begin{equation}
	\label{v_x_b}	
 V_b(y)= \frac{1}{ m v_F} \int\limits_{-\pi/2}^{3\pi/2} \frac{ d\alpha}{\pi} \, f_b(y,\alpha) \sin \alpha \: .
\end{equation}

The exact solution of the kinetic equation with the interparticle collision integral allows to calculate the near-edge ballistic contributions to the total current, $I$, and to the Hall voltage between the edges, $U_H$. However, such exact solution, apparently, can be performed only numerically, as, for example, it was done  in Ref.~\cite{Scaffidi2017,Raichev2022}.

In order to qualitatively describe the contributions from the near-edge layers to $I$ and $U_H$ by physically transparent analytical formulas, we propose a simplified model of the flow in this near-edge layer.

In this effective model first, we consider that there are a distinct bulk central layer and distinct near-edge semiballistic layers, divided by the virtually  interfaces at $y=\lambda$  and  $y=W-\lambda $ (see Fig.~\ref{Fig1}). The width of the semiballistic near-edge layers $ \lambda $ is the main parameter of this model. By its sense, it is of the order of the scattering length $ l_2  $, but smaller than $ l_2 $:  for example, $ \lambda  = l_2/2 $.

We assume that the electrons in the central bulk layer $\lambda < y < W- \lambda $ have purely hydrodynamic distribution function whose first and second angular harmonics are much greater than its higher harmonics:
  \begin{equation}
	\label{F_hydro}
\begin{array}{c}
\displaystyle
	f_h(y,\alpha)= m v_F  V_h(y) \sin\alpha \,  -
\\ \\
\displaystyle
- \,  2 \, V_{h}'(y)  \, [ \, \eta_{xx} \sin 2 \alpha + \eta_{xy} \cos 2\alpha \, ]
\: .
\end{array}
\end{equation}
The first harmonic of $f_h$ is proportional to the averaged electron velocity $V_h(y)$, which is the quasi-equilibrium hydrodynamic velocity in this region, and the second one is proportional to the non-equilibrium momentum flux components $\Pi_{xy} \sim V_h'$ and $\Pi_{yy} \sim V_h'$ .

As for near-edge regions, first, we suppose that the electrons incoming from the bulk region have the distribution function equal to the hydrodynamic one (\ref{F_hydro}) at the interfaces $y=\lambda, W- \lambda$:
\begin{equation}
\label{F_ballistic_bh_boundary_left}
  \begin{array}{l}
  \displaystyle
     f_{b,l}^{-}(\lambda,\alpha)=f_{h}(\lambda,\alpha)
        \, , \quad
        \frac{ \pi }{ 2 }  <\alpha < \frac{ 3\pi }{ 2}
         \:,
  \\\\
  \displaystyle
     f_{b,r}^{+}(W-\lambda,\alpha) = f_{h}(W-\lambda,\alpha)
          ,
          \; \; \; \;
         - \frac{\pi}{2} <\alpha <  \frac{ \pi }{ 2}
   ,
  \end{array}
\end{equation}
whereas the electrons reflected from the sample edges are described by the ballistic distributions isotropic with respect to velocity direction [see Eq.~(\ref{diff_b_c})].

Second, our model is based on the assumptions concerning the interparticle scattering in the near-edge layers $0<y<\lambda$ and $ W-\lambda < y < W $. From the one hand, the interparticle scattering of electrons with the angles~$\alpha$ being close to the sample direction:
\begin{equation}
\label{vic}
|\pi/2 - |\alpha| \, | <\delta
\:,
\end{equation}
is supposed to be infinitely fast, so that distribution function $f_{b,l/r}^{\pm} (y,\alpha)$ at such $y$ and $\alpha$ is zero (see Fig.~\ref{Fig1}). From the other hand, in the near-edge layer interparticle collisions are absent for the electrons with the velocities directed not too close to the edge directions:
\begin{equation}
  | \pi /2 -|\alpha |  | >\delta
  \:.
\end{equation}
Such model values of collision rates are a rough model of the fact that in near-edge layers electrons moving in directions close to the edge direction have enough time to collide with other electrons, while electrons moving in directions close to the normals to the edge do not have enough time to collide with other electrons in the near-edge layers.
The following estimate is used for the cut off parameter $\delta=\delta_{\lambda}$:
\begin{equation}
 \label{lambda_layers_sharp}
\delta  _ \lambda = \arcsin(\lambda/l_2)
 \:.
\end{equation}
Indeed, it is seen from Fig.~\ref{Fig1} that $\pi /2 - \delta$  is the maximal absolute value of the velocity angle $\alpha$, at which electron can move between the lines $ y = 0 $ and $ y = \lambda $ without collisions.

In this section we imply that the relaxation times of all angular harmonics of the distribution function in the near-edge layers are the same and equal to the relaxation time $\tau_2$ of the second harmonic. This simplifying assumption allows to propose  an even simpler model of the near-edge layers [see Fig.~\ref{Fig1}]. Namely, we consider that they have the widths $\lambda <l_2$ (say, $ \lambda = l_2/2$), in which a part of electrons with velocity angles $|\pi/2 -|\alpha|| >\delta $ do not scatter on other ones and either collide only with the edges or reach the hydrodynamic layer. The other part of electrons with the angles $ |\pi/2 -|\alpha|| < \delta $ is considered to be  scattered on all electrons in the layers  infinitely fast and give no contribution to macroscopic  quantities.

The solution of the ballistic kinetic equation~(\ref{kin}) with the boundary conditions (\ref{diff_b_c}), (\ref{diff_b_c2}), and (\ref{F_ballistic_bh_boundary_left})  in the near-edge layers $ 0<y<\lambda $ and  $ W- \lambda <y<W $, $| \pi /2 -| \alpha |  | >\delta $, yields  the distribution function $f_b(y,\alpha)$ which is expressed via the hydrodynamic function~(\ref{F_hydro}) at the boundaries $y = \lambda, W- \lambda$.

Third,   we also need to formulate some effective boundary conditions at  $y=\lambda,W - \lambda$ in order to find the function $V_h(y)$, which determines the hydrodynamic distribution function. For this purpose, we formulate the continuity conditions of the component of the momentum flow tensor $\Pi _{xy}(y)$:
\begin{equation}
  \label{ed_Pi}
  \begin{array}{l}
  \Pi_{ xy,l} (\lambda) = \Pi_{ xy,h} (\lambda)
  \, ,
 \\
   \Pi_{ xy,r} (W-\lambda) = \Pi_{ xy,h} (W-\lambda)
   \,.
   \end{array}
\end{equation}
Note that velocity $V (y)$ at $ y = \lambda $ and $ y =W -\lambda $ is discontinuous in the proposed model.   The values $\Pi_{xy} =   \Pi_{ xy,l} ,  \, \Pi_{ xy,r}  ,\,  \Pi_{ xy,h} $ are the components of the momentum flux tensor per one electron in the ballistic and in the hydrodynamic regions  (equal to the shear stress components with the factor ``-1''):
  \begin{equation}
  \label{Pi0}
\left[ \begin{array}{c}
\Pi_{xy}(y)
\\
\Pi_{yy}(y)
\end{array}\right]
=
\int \limits_{-\pi/2 }^{3\pi/2}  \frac{ d \alpha }{ 2 \pi } \,  f  (y,\alpha)
\left[ \begin{array}{c}
\sin 2 \alpha
\\
\cos 2 \alpha
\end{array}\right]
.
\end{equation}
Condition~(\ref{ed_Pi}) indicates that  there is no any extra relaxation or generation of tangential momentum of electrons~$p_x$ at
the virtual   interfaces $y=\lambda$ and $y=W-\lambda$.

As it was formulated above, the interparticle scattering in the near-edge layers in our model is supposed to be very fast for electron with velocities close to the edge directions, so as the non-equilibrium distribution function for such electrons  is supposed to be negligible. Therefore the vicinities~(\ref{vic}) of the angles $\alpha=\pm\pi/2$ are omitted when integrating any function $\Phi(\alpha ,y)$, for example, $f  (y,\alpha) \sin 2 \alpha $ and  $f  (y,\alpha) \cos 2 \alpha $  in Eqs.~(\ref{Pi0}) over $\alpha $ at $ y $ in the near-edge layers, $0<y<\lambda$ and $W-\lambda < y < W$.

\begin{figure}[t!]
	\includegraphics[width=0.85\linewidth]{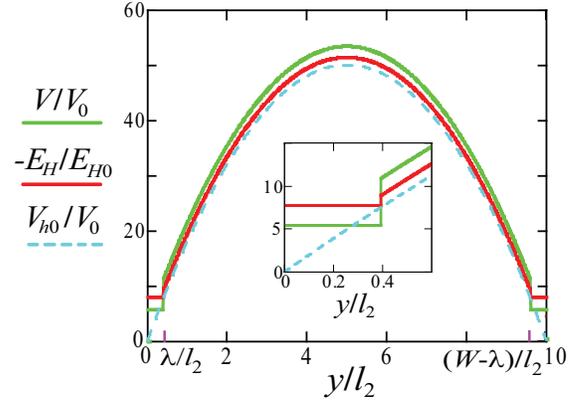}
	\caption{\label{Fig2} Profiles of the electron mean velocity $V(y)$ and the Hall electric field $E_H(y)$ calculated within ``sharp-layer'' model [equations (\ref{Velocity_ballistic0}), (\ref{Velocity_hydro0}), (\ref{EHb0}), and (\ref{E_o})].  The velocity is plotted in the unit $V_0 = eE_0\tau_2 /m $, while the Hall electric field is plotted un the unit $E_{H 0} = \omega_c \tau_2 E_0$. The width of the near-edge layers is taken to be qual to  $ \lambda = 0.39\, l_2 $, and the sample width is $W=10l_2$. The parameter of the refined boundary condition  is $\xi = 0.83 \, l_2  $. In the near-edge layers  the values $V(y)$ and $E_H(y)$  turn out to be homogeneous, unlike for the the purely hydrodynamic result for  $ V(y) $~(\ref{V_H_simplest})  and  $ E_H(y) $~(\ref{E_H_hydr}). Inset shows the vicinity of the left side of the main plot in a large scale.}
\end{figure}

In Appendix~B, we calculate the distribution function in zeroth order by magnetic field  in both the near-edge and the central bulk regions from an approximate solution of the formulated above equations (\ref{NS_eq}), (\ref{kin})-(\ref{diff_b_c2}), and (\ref{ed_Pi}). In particular,
the continuity condition of the momentum flux~(\ref{ed_Pi}) is used to obtain the boundary conditions for  the Navier-Stokes equation (\ref{NS_eq}) at the virtually interfaces $y=\lambda, W-\lambda$   in the central region $\lambda < y < W- \lambda $.   From the obtained distribution function we calculate the mean velocity $V(y)$.
It turns out to be flat in the near-edge layers:
\begin{equation}
	\label{Velocity_ballistic0}
	V_{b,l/r}=\frac{1}{2}\, V_h(\lambda) \Big[\,1-\frac{2\delta + \sin(2\delta )}{\pi}\,\Big]
   \:,
\end{equation}
and parabolic in the central region:
\begin{gather}
	\label{Velocity_hydro0}
\begin{array}{c}
   \displaystyle
	V_h(y)=\frac{2 e E_0}{mv_F^2 \tau_2}
     [ \, (W -y) y   \,+
         \\
    \\
    \displaystyle
     + \,
      (\xi -\lambda ) W + (\lambda -2\xi )\lambda \, ]
      \, .
      \end{array}
\end{gather}
Here the parameter $\delta =\delta _\lambda$  is given be Eq.~(\ref{lambda_layers_sharp}) and the parameter $\xi$,  which satisfies the relations  $\xi \sim l_2$, $\xi > \lambda $, is a function of $\delta$ calculated in Appendix~B:
\begin{equation}
   \label{xi_re0}
   \xi \, = \,  \frac{3\pi }{16  \,  \cos ^3 \delta } \, \Big[\,1 - \frac{ \sin (4\delta) -4\delta  }{2\pi} \, \Big]  \,  l_2
     \:.
\end{equation}
In Fig.~\ref{Fig2} we plot the  actual  [equations (\ref{Velocity_ballistic0}), (\ref{Velocity_hydro0})] and the purely Poiseuille [equation~(\ref{V_H_simplest})] velocity profiles.

The problem of calculation of the velocity in the bulk region can be reformulated via introducing the refined effective boundary conditions at the very sample edges $y = 0, \, W$. From Eqs.~(\ref{diff_b_c}), (\ref{diff_b_c2}), and~(\ref{ed_Pi}) with the proper distributioo we obtained (see Appendix~B) the boundary problem:
 \begin{equation}
\label{NS_and_ref_b_c_int}
\left\{
	\begin{matrix}
	\displaystyle
	\eta_{xx} \, (V_h^{e} )'' \, + \, e E_0/m \, = \, 0
   \\\\
   \displaystyle
		\left. \big[ \, l_{sl}\, (V_h^{e})'  \, \mp \,   V_h ^{e} \,\big] \right|_{y=0,W}=0
		\end{matrix}
\right.
   \:.
\end{equation}
Here $ l_{sl} = \xi  - \lambda $ is so called slipping length and $V_h^{e} $ is the effective velocity which coincides with the actual hydrodynamic velocity $V_h$ in the bulk region, but differ from in in the near-edge regions.

The  total current,  $    I = (e^2 n_0  / m) \int _0^W dy\, V(y)  $, corresponding to velocity profiles~(\ref{Velocity_ballistic0}), (\ref{Velocity_hydro0}) can be presented in the form $  I= I_h  + \Delta  I_b$, where:
\begin{equation}
	\label{I}
\begin{array}{c}
\displaystyle
    I_h = \frac{2e^2 n_0 E_0  }{mv_F^2\tau_2} \,
      \Big[
       \, \frac{W^3}{6} + W ^2 \,(\xi -\lambda ) +
            \\
\\
 \displaystyle
       + W \lambda\,(\lambda -2\xi )
     \, \Big],
     \\
\\
 \displaystyle
  \Delta   I_b =  \frac{ 2 e^2 n_0  E_0 }{ mv_F^2 \tau_2} \,   W \lambda \xi
   \Big[  \, 1- \frac{2\delta  + \sin(2\delta)}{\pi} \, \Big]
   \,.
\end{array}
\end{equation}
Here the contribution $I_b$ comes from the near-edge regions $0<y<\lambda$ and $W-\lambda < y < W$, while the contribution  $I_h$ comes from the central hydrodynamic region $\lambda <y <W$ and accounts for the flow in the near-edge regions via the refined boundary conditions~(\ref{xi}).
From Eq.~(\ref{I}) it is seen that the modification of $I_h$ due to  the refined boundary conditions in Eq.~(\ref{NS_and_ref_b_c_int}) is much greater than currents in the near-edge semiballistic layers $\Delta I _b$.

In Appendix~C we solve the formulated problem in first order by $\omega_c$ in order to study the Hall effect.
Namely, we use the obtained refined hydrodynamic velocity for  $V_h^e(y)$ (\ref{NS_and_ref_b_c_int}) and kinetic equation (\ref{kin}) in the near-edge regions in order to calculate the zero angular harmonic of the (generalized) distribution  function $f$~(\ref{f_tilda}):
\begin{equation}
 \label{0}
  f_{m=0} (y)=
  \int \limits_{-\pi/2}^{3\pi/2}
  \frac{d \alpha  }{2\pi}  \: f_b (y,\alpha )
\:,
\end{equation}
in the linear approximation by magnetic field: $f_{m=0}\sim \omega_c$. In Eq.~(\ref{0}) we also should omit the vicinities of the angles $\varphi = \pm \pi/2 $ in the near-edge layers. Such harmonic is approximately proportional to  the potential of the Hall electric field $E_H = - e \phi'$ in typical structures~\cite{Alekseev_Alekseeva_2019}:
  \begin{equation}
  f_{m=0} (y) \,\approx \,  e \phi (y)
  \:.
\end{equation}
 This relation reflects the relatively large magnitude of the electrostatic forces in the system compared to the hydrostatic forces associated with electron density gradients.
In the bulk region we solve the Navier-Stoke equation with the Hall velocity term.

The results of such calculation is as follows. The nontrivial contribution to the Hall effect  consists of two parts: the one from the Hall viscosity presented above in Eqs.~(\ref{E_H_hydr})-(\ref{Hall_re_corr_hydr}) and the one from the semiballistic flows in the near-edges regions $0 < y  < \lambda  $ and $0 < W- y   < \lambda   $. The Hall electric field $E_{H} ^b = \phi' (y)  =\mathrm{const}(y) $ corresponding to the calculated zero harmonic $f_{m=0}\sim \omega_c$~(\ref{0}) in the main order by the parameter $l_2/W$ takes the form:
\begin{gather}
\label{EHb0}
\begin{array}{r}
\displaystyle
	E_{H} ^b
 =\frac{ B}{c}
    \int  \limits  _{\frac{\pi}{2}+ \delta(\lambda) } ^{\frac{3\pi}{2} - \delta(\lambda) }
  \frac{d\alpha}{2\pi}
 \big[       
    l_2 V_h'(\lambda)  \frac{\cos2\alpha} {\cos\alpha} -V_h(\lambda)
  \big] .
\end{array}
\end{gather}
In the the bulk region we have the kinematic formula
\begin{equation}
\label{E_o}
  E_H^0 (y) = - (B/c) V_h(y)
\end{equation}
 with the refined velocity $V_h$~(\ref{Velocity_hydro0}) and add the Hall viscosity contribution according to Eq.~(\ref{E_H_hydr}).

In Fig.~\ref{Fig2} we draw the calculated Hall electric field in both the near-edge and the central regions.

\begin{figure}[t!]
	\includegraphics[width=0.7\linewidth]{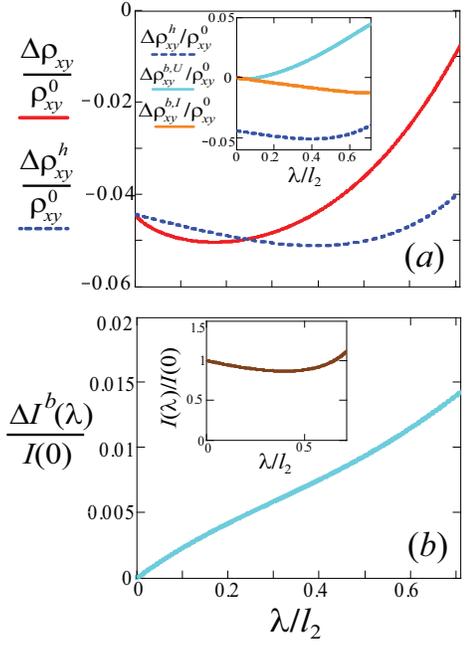}
	\caption{\label{Fig3} (a): Dependencies of the correction to the standard Hall resistance $ \varrho_{xy} ^ 0 = B / ( n_0  c ) $ on the width  of the near-edge layer $\lambda$, which is the free parameter of the developed analytical model. The sample width is $W=10  \, l_2$. The value $\Delta \varrho_{xy}$ is the total correction to $ \varrho_{xy} ^ 0 $, while $\Delta \varrho_{xy} ^h $ denotes the Hall viscosity correction calculated for the modified velocity profile $V_h(y)$ due to the refined boundary condition. Inset shows the three non-hydrodynamic corrections to the Hall effect defined according to the terms in  Eq.~(\ref{three}).  Namely, the value $\Delta \varrho_{xy} ^{b,U} $    is the contribution related to the Hall bias $U_{H}^h$ at the near-edge layers, while  $\Delta \varrho_{xy} ^{b,I} $ is the correction originating from the difference of the total current $I$ and the current in the bulk region $I_h$. Such difference proportional to the  current  $\Delta I _b$ in the near-edge layers. (b): Dependencies of   the current in the near-edge layers and the total current (in inset) on the width  of the near-edge layer $\lambda$. Note that the value $I(0)$ correspond to the purely hydrodynamic result~(\ref{II}).  }
\end{figure}

From Eqs.~(\ref{Velocity_hydro0}) and (\ref{EHb0}) it follows that the near-edge contribution  $ \Delta  U_{H,b} = 2 E_{H} ^b  \lambda $ to the Hall voltage $U_H = \phi(W) - \phi(0) $  takes the form:
\begin{equation}
\begin{array}{c}
 \displaystyle
    \Delta  U_{H} ^b = \frac{   2\omega_c  E_0   W \lambda  }{ l_2 }\,\Big\{
  \, \frac{\xi}{v_F} \, \Big[ \, 1- \frac{  2 \arcsin(\lambda ) }{\pi } \,  \Big]
 \, +
 \\
 \\
 \displaystyle
 + \,
  \frac{\tau_2}{ \pi }
 \Big[ \,
 4 r_\lambda
  +
  \ln
  \Big( \:
  \frac{ \displaystyle  1 -r_\lambda  }
  {\displaystyle  1 +r_\lambda }
  \: \Big)
 \, \Big]
 \Big\}
 \:,
\end{array}
\end{equation}
where $r_\lambda  =  \sqrt{1-\lambda^2 /l_2^2}$. The contribution to $U_H$ from the hydrodynamic region can be presented in the form:
\begin{gather}
\begin{array}{c}
\displaystyle
	U_{H} ^h  \, = \,\frac{ B  }{ec}\, \frac{ I_h  }{n_0} \, + \,    \Delta U_{H} ^h  \: ,
\\
\\
\displaystyle
\Delta U_{H} ^h    = \frac{m}{e^2 }\eta_{xy} [\, V_h' (W-\lambda) -V_h' (\lambda) \, ] =
\\
\\
\displaystyle
=
 -  2 \omega_c \tau_2 W  E_0 \:.
\end{array}
\end{gather}
Now we can express the total Hall voltage $U_H $ in the form containing the total current~$I$, the above  hydrodynamic correction~$\Delta U_{H,h} $, and the ballistic contributions~$ U_{H,b}$ and $ \Delta I _b$:
\begin{equation}
   U_H  \, = \,  \frac{B}{n_0ec} \,   I \,  +   \, \Delta U_{H} ^b  \,  + \,   \Delta U_{H} ^h  \,
 - \,    \frac{B}{n_0 e c}   \,  \Delta  I_b \:,
\end{equation}
where  the last term, $ \omega_c \Delta  I_b /n_0$, arises due to the difference of $I$ and $I_h$. The difference $\Delta  I_b = I-I_h$ is the near-edge semiballistic contribution is  determined by Eq.~~(\ref{I}).

The Hall resistivity in the zero and the first two orders by the  parameter $l_2/W\ll1$  and in the linear approximation by magnetic field $\omega _ c \tau_2 \ll 1 $ takes the form:
\begin{equation}
 \label{three}
  \rho_{xy}= \rho_{xy}^{0} + \Delta  \rho_{xy}^{h} + \Delta  \rho_{xy}^{b}
  \:,
\end{equation}
where $\rho_{xy}^{0} = B /(n_0ec)$ is the standard Hall resistance,  $ \Delta  \rho_{xy}^{h} $ is the bulk contribution from the Hall viscosity and $ \Delta  \rho_{xy}^{b} $  is the contribution from the semiballistic near-edge layers:
\begin{equation}
  \label{hall_resistivity_ball}
  \begin{array}{c}
  \displaystyle
	 \frac{ \Delta  \rho_{xy}^{h}  }{ \rho_{xy}^{0} } = -\frac{6 l_2^2}{W^2}
\:,
\\
\\
\displaystyle
	\frac{ \Delta  \rho_{xy}^{b} }{ \rho_{xy}^{0}} =
\, \frac{ 6  \lambda l _2  }{ \pi W^2} \,\Big[
\Big( \,4+
 \frac{  2\xi \lambda }{ l_2^2 } \,\Big) \, r _ \lambda   +
  \ln
  \Big(
  \frac{ \displaystyle  1 -r _ \lambda }
  {\displaystyle  1 + r _ \lambda }
   \Big)
  \Big]
  .
\end{array}
\end{equation}
We see from Eq.~(\ref{hall_resistivity_ball}) that  the  semiballistic contribution $\Delta  \rho_{xy}^{b}$ to the resistance  $\varrho_{xy} $ is of the same order of magnitude by the small parameter  $l_2/W \ll 1$ as the Hall viscosity contribution $\Delta  \rho_{xy}^{h}$. Herewith only $\Delta  \rho_{xy}^{b}$ depends on the model parameter $\lambda$ [via the  values $\delta = \delta _\lambda$, $\xi = \xi(\lambda)$, $r_\lambda$ related with $\lambda$].

In~Fig.~\ref{Fig3} we plot the dependencies of the hydrodynamic correction to the Hall resistance $ \Delta \rho_{xy}^{h}  $ and of the near-edge semiballistic contributions, the total one   $ \Delta \rho_{xy}^{b} $ and its components defined according to Eq.~(\ref{three}), on the near-edge layer width $ \lambda $. It is seen they are of the same order of magnitude, however strongly depend on the model parameter $\lambda$. This indicates the model provide a correct value of
$ \Delta \rho_{xy}^{h}  $ and $ \Delta \rho_{xy}^{b} $
by the order of magnitude. In that figure we also present the dependencies of the total current $I$ and the near-edge contribution $\Delta I _ b $ on $ \lambda $. The main part of the first  dependence  $I(\lambda)$ is related to the effect of the modified boundary conditions, however the value of the second contribution $\Delta I _ b $ is much smaller than the effect from the modification of the boundary conditions~(\ref{NS_and_ref_b_c_int}) on the velocity in the bulk.


\section{ Three-harmonic approximation for  distribution function of Poiseuille flow }
 Now  we develop another approach for the description of the near-edge semiballistic layers in a Poiseuille flow. This model is complementary to the model of the previous section in the following sense. Instead of making any simplification for the real-space structure of the flow, here we focus on the Fourier decomposition of the distribution function by the velocity angle~$ \alpha $.
We will derive the hydrodynamic-like equations taking into account the lowerest angular harmonics of distribution function up to the third one with the numbers $m=0,\pm1,\pm2,\pm3$.

Namely, we drop a simplified assumption, made in previous sections, that there are near-ballistic layers with sharp boundaries at $y = \lambda, \, W-\lambda$ and solve the kinetic equation with a non-zero collision integral:
\begin{equation}
\label{kin3}
	v_F  \cos\alpha  \, \frac{\partial \tilde{f}}{\partial y} - \omega_c \,  \frac{\partial \tilde{f}}{\partial \alpha} -  E_0 v_F \sin\alpha
       \, = \,
   \mathrm{St}[\tilde{f}] \,  ,
\end{equation}
in the whole bulk of the sample  with exact space dependencies of all values in the whole  sample, $0<y<W$,

Our key  model simplification is now that we use
a truncated distribution function $f$ whose Fourier decomposition by~$ \alpha $ contains only the harmonics of the zeroth and first three orders:
\begin{equation}
\label{F_hydro_3}
\begin{array}{c}
   \displaystyle
    \tilde{f}(y,\alpha) \, = \, -\,   f_{0}(y) \, +
   \\
   \displaystyle
 \,    + \sum\limits_{ m = 1 }^{3} [\, f_{mc}(y) \, \cos ( m \alpha) + f_{ms}(y) \, \sin ( m \alpha)  \, ] \: .
\end{array}
\end{equation}
Such distribution function $f$ takes into account the next order harmonics by~$ \alpha $, the third ones $\cos 3 \alpha , \, \sin 3 \alpha $,  beyond the hydrodynamic approximation~(\ref{F_hydro}). Therefore,  such $f$ approximately takes into account the ballistic contributions to the current and the Hall electric field from  the near-edge regions.  In other words, we consider that, by some reasons, the relaxation times of the fourth and higher harmonics, $\tau_4$, $\tau_5$, and so on, are much shorter than the times $\tau_2$ and $\tau_3$.

Below, as in the previous section, we will omit the tilde symbol in  $\tilde{f}$ for brevity.

The integral of inter-electronic collisions can be taken in the form:
\begin{equation}
   \mathrm{St}[\, f \, ] \,
   =
    \,  - \,
    \frac{  \hat{P}_2 [ \, f \, ]}{\tau_2}
   \, - \,
   \frac{  \hat{P}_3 [ \, f \, ]}{\tau_3}
\:,
 \end{equation}
where $\hat{P}_m$ are the projector operators on the $m$-angular harmonics subspace,  $\sin (m\alpha)$ and $\cos(m\alpha)$. Within this model, we can assume that the relaxation times $\tau_2$ and $\tau_3$  can have the same or different  orders of magnitudes.

For example, the last case is realized for  the degenerate Fermi gas or Fermi liquid  of electrons with a quadratic energy spectrum and the Coulomb interaction potential in samples with no disorder (see, for example, Ref.~\cite{Alekseev_Dmitriev_2020}).  The large difference in relaxation times  $\tau_2$ and $\tau_3$  in this case is due to the kinematic restrictions from the laws of conservation of energy and momentum and the singular dependence of the interaction potential on the momenta of the colliding electrons.
The first case may be realized for two-dimensional electrons with a non-quadratic spectrum (for example, in graphene) in structures with a gate, where the effective potential of the interparticle  interaction is short-range and there are no kinematic restrictions associated with the law of energy and momentum conservation (owing to the lack of proportionality between momentum and velocity).

It is shown in Appendix D that from the above kinetic equation for the distribution function, taking into account only the first three harmonics, are derived the closed system of differential equations (\ref{two}) for the velocity $V(y)$ and momentum flux components  $\Pi_{xy,yy}(y)$. In addition, it is shown there that some proper boundary conditions for $V(y)$ and  $\Pi_{xy,yy} (y)$ can be derived from the exact diffusive boundary conditions (\ref{diff_b_c}) and (\ref{diff_b_c2}) on the base of requirement of minimizing some residual simplest functional, the quadratic one. Such functional characterizes the deviation of the true distribution function that satisfies Eqs.~(\ref{diff_b_c}) and (\ref{diff_b_c2}) from the approximate distribution function (\ref{F_hydro_3}) and corresponding the the profiles  $V(y)$ and   $\Pi_{xy,yy}$ satisfying the obtained differential equations for $V$ and $\Pi_{xy,yy}$.

In order to find the corrections to the mean  velocity $V(y)$  and the  Hall electric field $E_H(y)$ from near-edge semiballistic layers,
 let us solve the formulated  equations (\ref{two}),  (\ref{conditions_3_res})-(\ref{conditions_3_res2}) for the regime  of weak magnetic fields, $\omega_c\tau_{2,3} \ll 1$, considered above for the sharp-layers model. It was shown  that in the zeroth and the first orders by the small parameters $\beta_2 = 2 \omega_c \tau_2 $ and $\beta_3 = 3 \omega_c \tau_3$ equations~(\ref{two}) become decoupled and take the simple form:
 \begin{equation}
 \label{two_p}
  \left\{
  	\begin{array}{l}
 \displaystyle
\eta_0  \, V'' \,= \,- \, E_0
\\\\
\displaystyle
 \lambda ^2  \, \Pi_{yy}'' \,= \,  \Pi_{yy}   + \eta_0 \beta_2  \, V'
\end{array}
\right.
.
\end{equation}
Here
\begin{equation} \lambda   =  \sqrt{l_2 l_3 } / 2 \end{equation}
 is the the characteristic length,  for which we intensionally use the same notation  $\lambda $ as for the width of the near-edge sharp layers in the model developed in the previous section, as the length $\lambda$ introduced here has the same meaning of the width of the near-edge layers where the ballistic contributions to the electron distributions are substantial. The boundary conditions~(\ref{conditions_3_res})-(\ref{conditions_3_res2}) in the limit $\beta_2,\, \beta_3 \ll1$ take the form:
\begin{equation}
\label{conditions_3_res_p}
\begin{array}{l}
\displaystyle
 \left.   \Big( \, V  \mp \frac{4}{3\pi} \, l_2 \,  V' \, \Big)  \right| _{y=0,W} =0
    \, ,
    \\
    \\
\displaystyle
 \left.   \Big[  \, \Pi_{yy} \mp  \frac{ 64 }{ 55 \pi    }  \, l_3\,  ( \Pi_{yy} '  + \beta_3 \frac{e E_0}{m} \Big]  \right| _{y=0,W} =0
   \: .
\end{array}
\end{equation}

The first rows of Eqs.~(\ref{two_p}) and (\ref{conditions_3_res_p}) are a closed system of equations for the velocity profile $V(y)$ in the zeroth order by magnetic field. They are similar to the Navier-Stokes equation~(\ref{NS_and_ref_b_c_int}) for the effective velocity $V_h^e(y)$ with refined boundary conditions~(\ref{NS_and_ref_b_c_int})  from the approach based on consideration of semi-ballistic near-edge layers. In the current approach the slipping length takes the form:
\begin{equation}
l_{sl}\, = \,  \frac{ 4}{ 3 \pi }\,  l_2
\end{equation}
[see the first of Eqs.~(\ref{conditions_3_res_p})]. The result  for the velocity is:
\begin{equation}
\label{vx_in}
\begin{array}{c}
 \displaystyle
    V(y) =\frac{eE_0}{ 2 m \eta_0} \, [\,
  y(W-y)
\,+\,
  l_{sl}W
\, ]
 \:.
\end{array}
\end{equation}
This velocity profile yields the current:
 \begin{equation}
\label{I_vx_in}
I= \frac{e^2 n_0E_0  }{ 2 \eta_0} \, \Big[ \,
  \frac{W^3}{6}
\,+\,
  l_{sl}W^2
\, \Big] \,.
 \end{equation}
We see that the  accounting for the third harmonics in distribution~(\ref{F_hydro_3}) and corresponding boundary conditions~(\ref{conditions_3_res_p}) leads to the decrease of the rate of momentum relaxation on the edges [as compared to the purely hydrodynamic solution~$V(y)$~(\ref{V_H_simplest}) obtained from  the simplest conditions $V_{y=0,\,W} = 0 $ which overestimate the relaxation rate of the electron scattering on the edges] and, thus, to the increase of the flow magnitude $V(y)$ and $I$ [as compared to Eq.~(\ref{V_H_simplest_lim})].

Using  the second rows of Eqs.~(\ref{two_p}) and (\ref{conditions_3_res_p}), we calculate  the distribution of the momentum flux component $\Pi_{yy}(y)$. In the limit of  wide samples, $W \gg l_2$, we obtain:
\begin{equation}
\label{PI_xx}
 \Pi_{yy} (y)
=
 \beta _2 eE_{0}
 \Big[  \Big( y - \frac{W}{2} \Big)
 \, - \,    \frac{W}{2}  \,   \frac{e^{-\frac{W-y}{\lambda}}  -    e^{-\frac{y}{\lambda}} }{1+ \tilde{l }_{sc} /\lambda }
 \Big].
\end{equation}
where we introduced the ``second slipping length'':
\begin{equation}
  \tilde{l }_{sc} \, = \, \frac{64}{ 55\pi} \, l_3
  \:.
\end{equation}
It   comes from the second of boundary conditions~(\ref{conditions_3_res_p}).

\begin{figure}[t!]
	\includegraphics[width=0.7\linewidth]{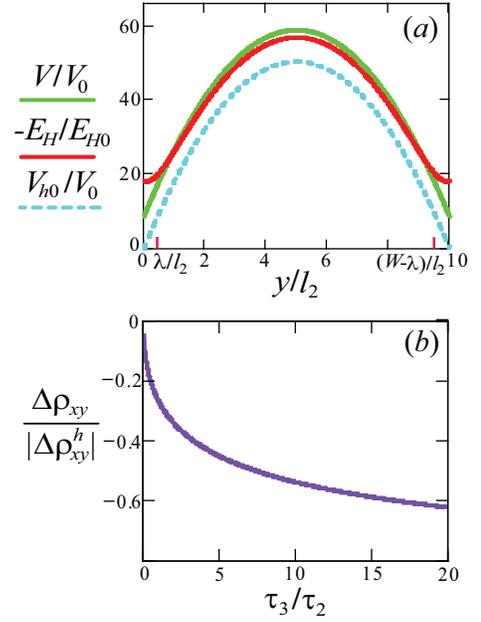}
	\caption{\label{Fig4} (a): The velocity profile $V(y)$ and the Hall electric field $E_H (y)$ calculated within the second proposed model based on the three-harmonic approximation for the distribution function. The velocity is plotted in the unit $V_0 = eE_0\tau_2 /m $, while the Hall electric field is plotted in the unit $E_{H 0} = \omega_c \tau_2 E_0$. The sample width is $W= 10 l_2$, and the scattering times $\tau_2$ and $\tau_3$ are equal one to another. For comparison, dashed line shows the parabolic profile of the Poiseuille flow $V_{h0}(y)$ calculated with the no-slip boundary conditions  $V_{h0}(y=\pm W/2) = 0$. (b): The dependence of the  correction to the standard Hall resistance $\varrho_{xy}^0$, divided on the purely hydrodynamic correction,  on the ratio of the relaxation rates of the third and the second angular harmonics in the limit of a wide sample: $W \gg l_2$ and $W \gg l_3$. }
\end{figure}

From the first of Eqs.~(\ref{total_eq_4_functions}),
\begin{equation}
eE_H = -m\omega_c V +\Pi_{yy}'
\,,
\end{equation}
 and Eq.~(\ref{PI_xx}) we obtain expressions for the Hall electric field:
\begin{equation}
\label{hall_e}
\begin{array}{c}
\displaystyle
E_{H}  (y) \, = \,
  - 2\omega_c E_0\,   \Big[\, \frac {y(W-y)+\xi W }{v_F^2 \tau_2}   \, -
  \\\\
  \displaystyle
  - \, \tau_2\,
 + \frac{W\tau_2}{2\lambda}   \, \frac{e^{-\frac{W-y}{\lambda}}  +  \,  e^{-\frac{y}{\lambda}} }{1+ \tilde{l }_{sc}/\lambda }
 \, \Big] .
\end{array}
\end{equation}
Here the first term is the Hall viscosity contribution and  the second one is the near-edge contribution.

For the Hall voltage we have:
\begin{equation}
\label{hall_voltage}
\begin{array}{c}
\displaystyle
 U_{H} =
\dfrac{m\omega_c }{e^2 n_{0}c} \, I  - \dfrac{ \beta_2 E_{0} W  }{ 1+\lambda / \tilde{l }_{sc}}
  \: .
\end{array}
\end{equation}
Note that one should keep in mind that this result is derived in the limit $\omega_c \tau_{2,3} \ll 1$, thus the the cases $\tau_2 \ll \tau_3$ or visa versa, $\tau_2 \gg \tau_3$, can be considered only to a limited extent.

Finally, for the Hall resistance we obtain:
\begin{equation}
\label{hall_res_3h}
\varrho_{xy} = \varrho_{xy} ^0 + \Delta \varrho_{xy}
\, , \quad \Delta \varrho_{xy}  = \dfrac{ \Delta \varrho_{xy} ^h  }{ 1+\lambda / \tilde{l}_{sc} }    \: .
\end{equation}
We remind that the purely hydrodynamic correction  $ \Delta \varrho_{xy} ^h $~(\ref{Hall_re_corr_hydr}) is negative.

In Fig.~\ref{Fig4}(a) we plot the profiles  $V(y)$ and $E_H (y)$ for a sufficiently wide sample: $W\gg l_2$.

First, it is seen that  the refined boundary conditions~(\ref{conditions_3_res_p}) for the velocity $V(y)$ induces the increase of $V(y)$ in the whole sample as compared with the solution of the Navier-Stokes equation $V_{h0}(y)$  with the simplest boundary conditions $V_{h0}|_{y=0,\, W} =0 $. Indeed, in the last conditions the rate of the diffusive scattering on the edges is overestimated, therefore accounting of the refined boundary conditions leads increase of the velocity.

Second, one can see at Fig.~\ref{Fig4}(a)  that the correction  to the Hall field $E_H(y) $ from the Hall viscosity in the bulk, given by the term $  2 \omega_c \tau_2  E_0 $ in  result~(\ref{hall_e}), has the sign being opposite to the sign of the main ``kinematic'' part of the Hall field, $E_{H}^{kin}(y)  = -BV(y)/c$, in which the velocity $V(y)$ is calculated within the three-harmonic approximation.

Third, it is noteworthy that in the near-edge layers the value $E_H(y)$, mainly determined by the last term in Eq.~(\ref{hall_e}), describes the mixed ballistic-hydrodynamic flow in the near-edge layers. Such contribution has the same sign as the main contribution in the bulk, $ - E_{H}^{kin} (y) = BV(y)/c$ [see Fig.~\ref{Fig4}(a)].  Such behavior of $E_H(y)$ in the near-edge layers leads to the decrease of the absolute value of the total correction $\Delta \varrho_{xy}$ to the Hall resistance as compared with the the Hall viscosity contribution  $ \Delta \varrho_{xy} ^h $   [see Eq. (\ref{hall_res_3h}) and Fig.~\ref{Fig4}(b)]. However, the sign of the total correction $\Delta \varrho_{xy}$ remains  negative, as like the sign of $\Delta \varrho_{xy}^h$.

In Fig.~\ref{Fig4}(b) we present the dependence of the total correction  $ \Delta \varrho_{xy}  $ on the ratio of the relaxation times $\tau_2$ and $\tau_3$. It is seen that the larger is the ratio $\tau_3 / \tau_2$,  the larger is the absolute value of the total correction   $ \Delta \varrho_{xy}  $. Note that one cannot consider the limits $ \tau_{2} \to \infty $ or $ \tau_{3} \to \infty $  within the derived resulting formulas since   we imply only the limit of the weak magnetic fields, when  $\omega_c \tau_{2}  \ll 1$ and  $\omega_c \tau_{3}  \ll 1$.

In this way, we demonstrated that the value   $ \Delta \varrho_{xy}  $ is determined by both the bulk and the near-edge contributions. Herewith the shape of the near-edge layer in the three-harmonic model is controlled  by both the times $\tau_2$ and $\tau_3$. The resulting value $ \Delta \varrho_{xy}  $  is close to the hydrodynamic limit $ \Delta \varrho_{xy}^h  $ not in the case $\tau_3 / \tau_2 \ll 1 $, when the near-edge layers are narrow, but in the case  $\tau_3 / \tau_2 \gg 1 $. In the last case these layers are wide, but the large ``second slipping length''  in Eq.~(\ref{hall_e}), $\tilde{l}_{sc} \gg \lambda $, eliminates the contribution from wide near-edge layers.

The neglected high-order harmonics in the distribution function, apparently, may  lead to contributions to the current density and the Hall field from the near-edge layers $0<y \lesssim l_{m}$ and $ 0<W-y  \lesssim l_{m}$ of the same order of magnitude as the contributions associated with $f_{3c}$ and $f_{3s}$. At least, this seems to be true when all relaxation times $\tau_m$ are of the same order of magnitude. In the last case, the resulting values of the current and the Hall resistance within the three-harmonic approximation are calculated, apparently,  up to  factors of the order of unity.

\section{ Conclusion}
We have constructed  the two models of the  near-edge layers of a Poiseuille flow of two-dimensional electrons. First, we have derived within them the refined boundary conditions for the hydrodynamic velocity $V(y)$ containing the  slipping length~$l_{sl} \sim l_2$. This already known result (see, for example, Ref.~\cite{Kiselev2019,Raichev2022}) evidences the relevance of the proposed models. The increase of~$V(y)$ in the bulk as well as near the edges relative to the solution $V_{h0}(y)$ with the simplest  conditions $V_{h0}|_{y=0,W }  =0 $ appears due to a more precise accounting of the scattering on the edges by the refined boundary conditions (\ref{NS_and_ref_b_c_int}).

Second, we have studied the Hall effect in a Poiseuille flow.  The two proposed models yield  qualitatively the same results for the Hall electric field in the near-edge layers due to formation of the semiballistic flow there. Such field leads to the contribution  in the sample Hall resistance  of the same order of magnitude as the already-known  bulk  contribution from the Hall viscosity. This result can be presented the form:
\begin{equation}
\label{l}
 \begin{array}{c}
 \displaystyle
 \varrho_{xy} \,  = \, \varrho_{xy} ^0 \,  + \, \Delta    \varrho_{xy}  ^h \, + \,  \Delta    \varrho_{xy}  ^b\:,
 \\\\
 \displaystyle
   \frac{\Delta    \varrho_{xy}  ^{\nu} }{\varrho_{xy} ^0}  \, = \,  C_{\nu}\, \frac{l_2^2 }{W^2 }
   \:,\qquad
   \nu= h,b
\:,
\end{array}
\end{equation}
where the numbers $ C_{\nu} $ corresponds to the hydrodynamic and the semiballistic contributions, $\nu \, = \, h , \,  b $. They  depend on the particular parameter of the model  and typically are of the order of unity [see Eqs.~(\ref{hall_resistivity_ball}),~(\ref{hall_res_3h}) and  Figs.~3(a),~4(b)].

We believe   that our consideration has revealed the physical essence of the near-edge contribution to the Hall effect in a Poiseuille flow of the electron  fluid.
The comparison of the results of two our models one with another evidences that  the magnitude of the  near-edge contribution to the Hall resistance  $\varrho_{xy}$  is reliably calculated within them up to a numeric coefficient.
Moreover, this comparison hints that the actual value of the near-edge contribution strongly depends on the  structure of the sample edges as well as on the particular character of the relaxation processes in the bulk of a sample (for example, on the ratio of the relaxation times  $\tau_2/\tau_3$).

The exact value of the near-edge contribution in  $\varrho_{xy}$, apparently, can be calculated only by a numerical solution of the kinetic equation, similar  to the one performed in Ref.~\cite{Scaffidi2017}. In that publication it was established that the non-trivial part  of the Hall resistance, $\Delta \varrho_{xy} = \varrho_{xy} -\varrho_{xy} ^0 $, is negative.  This sign is consistent with  our results~(\ref{hall_resistivity_ball}) and~(\ref{hall_res_3h}).

Based on these theoretical results, first, we argue that the Hall voltage and the Hall resistance measured  by usual techniques by metallic contacts attached to the sample edges should  contain the Hall viscosity contribution as well as the comparable contribution from the near-edge semiballistic regions. Second, it is seen from performed consideration  that the Hall viscosity $\eta_{xy}$ determines the correction to the Hall field $E_H(y)$ in the bulk of a viscous flow. Thus this kinetic coefficient  may be measured separately by some contactless techniques  similar to ones used in Refs.~\cite{Sulpizio2019,Ku2020}.

\section{ Acknowledgments}

We thank M.~I.~Dyakonov and I.~V.~Gornyi for fruitful discussions.

The study was supported by the Russian Foundation for Basic Research (Grant No. 19-02-00999).

\appendix

\section{Characteristics of the Poiseuille flow at low magnetic fields  within purely hydrodynamic model }
In this section we present the formulas for the flow characteristics in the purely hydrodynamic model (Section II in the main text)
for the limit of the weak magnetic field,  $\omega_c \tau_2 \ll 1 $.

 In the zero and the first orders by $\omega_c $ the viscosity coefficients takes the form:   $	\eta_{xx}= v_F^2\tau_2/4 $ and $ \eta_{xy} = v_F^2 \omega_c  \tau_2^2  /2 $. General equations~(\ref{V_H_simplest})-(\ref{U_H}) for the hydrodynamical velocity, the current, the Hall electric field, and the Hall voltage turns into:
\begin{equation}
 \label{V_H_simplest_lim}
    V_h(y) =
      \frac{ 2 e E_0 }{ m v_F^2 \tau_2 }  \, y \, (W-y)
  \:, \quad
      I=  \frac{ e^2 n_0 E_0 W^3 }{3 m v_F^2 \tau_{2} }
    \:,
\end{equation}
\begin{equation}
\label{E_H_hydr_lim}
   E_H ^h (y) = - 2 \omega_c E_0 \,
   \Big[
     \frac{ y \, (W-y) }{ v_F^2 \tau_2}  -   \tau_2
    \Big]
 \:
  \:,
\end{equation}
\begin{equation}
\label{U_H_lim}
   U_H ^h  \, =  \,  \omega_c   E_0   W   \Big[ \frac{W^2}{3 v_F^2 \tau_2 } - 2  \tau_2  \Big]
   \: .
\end{equation}
These expressions, in particular, yield result~(\ref{Hall_re_corr_hydr}) for the purely hydrodynamic correction to the Hall resistance due to the Hall viscosity.

\section{Flow characteristics in zeroth order by $\omega_c$ and refined boundary conditions within ``sharp-edge'' model}
In this section, we calculate the distribution function in the zeroth order by magnetic field. Such function leads to obtaining of the refined effective boundary conditions~(\ref{NS_and_ref_b_c_int}) for the velocity profile~$V(y)$. The corresponding solution of the Navier-Stokes equation accounts for the correction to the Poiseuille velocity profile~(\ref{V_H_simplest_lim}) from  exact character of the electron dynamics near the rough sample edges.

The hydrodynamic distribution function~(\ref{F_hydro}) in the zeroth order by magnetic field takes the form:
\begin{equation}
 \label{F_hydro0}
	f_h^0(y,\alpha)= m v_F V_h(y) \sin \alpha - \frac{ m v_F^2 \tau_2 }{2}  V_h'(y) \sin 2\alpha.
\end{equation}
Ballistic distribution functions $f_{l}^{0}= f_{b,l}|_{\omega_c=0}$ and $f_{r}^{0}=f_{b,r} |_{\omega_c=0} $ in the zeroth order by $\omega_c$  are  calculated from the truncated kinetic equations:
\begin{equation}
\label{0_order_ball_eqs}
\cos \alpha  \, \frac{\partial f_{l/r}^{0}}{\partial y} - e E_0 \, \sin \alpha =0 \:,
\end{equation}
 with the boundary conditions:
\begin{equation}
 \label{bc_zero_order__lambda}
	\begin{array}{l}		
 \displaystyle
\left. \big( \, f_{l}^{0,-} \, = \, f_h^{0} \, \big) \, \right|
  _   { \,  y = \lambda  \,  ,  \; \frac{ \pi}{2} <\alpha < \frac{ 3 \pi}{2} }
  \, ,
\\\\
\displaystyle
\left.	f_{l}^{0,+} \, \right|_   { \,  y = 0 \,  ,  \;
    - \frac{\pi}{2} <\alpha < \frac{ \pi}{2} \, } =0
      \, ,
       \end{array}
\end{equation}
and
\begin{equation}
\label{bc_zero_order__Wlambda}
	\begin{array}{l}
 \displaystyle
	\left.	\big(\, f_{r}^{0,+} \, = \, f_h^{0} \,\big)  \, \right|_   { \,  y = W-\lambda \,  ,  \;
 - \frac{ \pi}{2} <\alpha < \frac{ \pi}{2}}
\,  ,
   \\\\
 \displaystyle
	\left.	f_{r}^{0,-}  \, \right| _   {
\,  y = \lambda  \,  ,  \;
 \frac{ \pi}{2} <\alpha < \frac{ 3 \pi}{2}
} =0
 \:.
   \end{array}
\end{equation}
Boundary conditions for the electrons reflected from the edges [second rows of Eqs.~(\ref{bc_zero_order__lambda}) and (\ref{bc_zero_order__Wlambda})] are the consequence of the abscence of the incoming transverse non-equilibrium current flow throught the interfaces from the hydrodynamic region at zero magnetic field [see Eq.~(\ref{F_hydro0})]. The result for the left-edge layer function $	f_{l}^{0}$ takes the form:
\begin{equation}
 \label{f_b_0_results}
\begin{array}{c}
\displaystyle	f_{l}^{0,-}(y,\alpha)= m v_F V_h(\lambda) \, \sin\alpha \, -
\\\\
\displaystyle
- \,  \frac{ m v_F^2 \tau _2 }{2} \, V_h'(\lambda)  \,  \sin2\alpha +  e E_0  \, (y - \lambda ) \, {\rm tg} \,\alpha
\end{array}
\end{equation}
and
\begin{equation}
 \label{f_b_0_results2}
	f_{l}^{0,+}(y,\alpha) = e E_0 \, y\,  {\rm tg}\,\alpha
 \:.
\end{equation}
The right-edge function $f_{r}^{0}$ is obtained from  $f_{l}^{0}$  by substitutions $y \to W-y$ and $\alpha \to \pi - \alpha $.  For the actual $V(y)$ given by equation~(\ref{V_H_simplest}) or by its more precise variant (see below) the purely ballistic terms in Eqs.~(\ref{f_b_0_results}) and (\ref{f_b_0_results2}), proportional to $ {\rm tg} \,\alpha$, have the smaller order of magnitude by the parameter $l_2/W$  than the other terms originating from boundary conditions~(\ref{bc_zero_order__lambda}).

The momentum flux components (\ref{Pi0}) corresponding  to the functions $f_{l}^{0}$~(\ref{f_b_0_results}),~(\ref{f_b_0_results2})  is given by the formulas accounting the absence of electrons with the velocities being close to the edge direction:
  \begin{equation}
  \label{Pi2}
 \begin{array}{c}
\Pi_{xy}(y)
\\
\Pi_{yy}(y)
\end{array}
\Big\}
 =  \Big[
\int \limits_{-\frac{ \pi}{2} +\delta }^{ \frac{  \pi}{2} - \delta  }
  + \int \limits_{ \frac{  \pi}{2}+ \delta  }^{ \frac{  3\pi}{2} - \delta  }
\Big]
\,
 \frac{d \alpha  }{ 2 \pi }  \,  f _b  (y,\alpha)
 \Big\{ \begin{array}{c}
\sin 2 \alpha
\\
\cos 2 \alpha
\end{array}
.
\end{equation}
A calculation by this formula with the distribution functions
(\ref{F_hydro0}), (\ref{f_b_0_results}), and (\ref{f_b_0_results2})
 in the main order by  $l_2/W$ yields:
	$ \Pi_{yy,h}^{0}(y)=\Pi_{yy,b}^{0}(y)=0 $,
\begin{equation}
  \label{zero_boundary}
	\Pi_{xy,h}^{0}(y)=- \frac{mv_F^2 \tau_2 }{4} \, V_h'(y) \:,
\end{equation}
and
\begin{equation}
\begin{array}{c}
\displaystyle	
\Pi_{xy,l}^{0}(y)
     = - \frac{2mv_F}{3\pi} V_h(\lambda) \, \cos^3 \delta
   \, -
   \\
   \\
   \displaystyle
     -\, \frac{ m v_F^2 \tau_2}{ 8  } \, V_h'(\lambda) \, \Big[\,1 + \frac{ \sin (4\delta) -4\delta  }{2\pi} \, \Big] \,,
     \end{array}
\end{equation}
and similarly for $\Pi_{xy,r}^{0}$. Substitution of these formulas for  $\Pi_{xy,h}^{0}$ and $\Pi_{xy, l/r}^{0}$  into the boundary conditions~(\ref{ed_Pi}) at $y=\lambda, \, W-\lambda$ results in the refined boundary conditions for $ V (y) $.  The values of $V$ and $V'$ near the edges are connected by the length parameter $\xi = \xi(\delta)$:
\begin{equation}
\label{xi}
\begin{array}{c}
\displaystyle
V |_{y=\lambda }
 =
  \xi \, V' |_{y = \lambda }
  \:
 ,
\\
\\
\displaystyle
 V |_{y= W - \lambda}
 =
   - \xi \, V' |_{y = W - \lambda}  \: ,
\end{array}
\end{equation}
where
\begin{equation}
   \label{xi_re}
   \xi \, = \,  \frac{3\pi }{16  \,  \cos ^3 \delta } \, \Big[\,1 - \frac{ \sin (4\delta) -4\delta  }{2\pi} \, \Big]  \,  l_2
     \:,
\end{equation}
and $\delta =\delta  _\lambda $ is given by Eq.~(\ref{lambda_layers_sharp}).

We note that our consideration, based on the introduced interfaces at  $ y=\lambda , \, W - \lambda $, is a simplified variant of the exact derivation of boundary condition~(\ref{xi}) from a solution of the kinetic equation of the type of Eq.~(\ref{kin}) taking into account  the interparticle collision integral everywhere  in the near-edge regions as well as the exact boundary conditions~(\ref{diff_b_c})  and~(\ref{diff_b_c2})   at the very sample edges $y=0,W$. In Refs.~\cite{Kiselev2019,Raichev2022} similar derivation of the effective boundary conditions, analogous to Eqs.~(\ref{xi}) was performed, based on the approximate solution of the kinetic equation in the near-edge layers.

Now the problem of finding the function $V_h(y)$ in the hydrodynamic region $\lambda < y < W-\lambda$ in the main order by the small parameter~$l_2/W  \ll 1$ can be formulated in the form
\begin{equation}
\label{NS_and_ref_b_c}
\left\{
	\begin{matrix}
	\displaystyle
	\eta_{xx} \, V_h'' \, + \,  e E_0/m \, = \, 0
   \\\\
   \displaystyle
		\left. \big( \, \xi \, V_h'  \, \mp \,   V_h  \,\big) \right|_{y=\lambda,W-\lambda}=0
		\end{matrix}
\right.
\:.
\end{equation}
The solution of this problem up to the third order by~$l_2 /W $ is:
\begin{gather}
	\label{Velocity_hydro}
\begin{array}{c}
   \displaystyle
	V_h(y)=\frac{2 e E_0}{mv_F^2 \tau_2}
     [ \, (W -y) y +
         \\
    \\
    \displaystyle
     +
      (\xi -\lambda ) W + (\lambda -2\xi )\lambda \, ].
      \end{array}
\end{gather}
The velocity and its derivative at the edges of the hydrodynamic region  $y=\lambda ,W-\lambda $, which enter ballistic distribution~$f_b$~(\ref{f_b_0_results}),  in the main order by $l_2 /W $ are:
\begin{equation}
 \label{VV'}
\begin{array}{c}
   \displaystyle
   V_h(\lambda ) = \frac{ 2e E_0 W \xi  }{mv_F^2 \tau_2}
    \:, \qquad
    V'_h(\lambda ) = \pm \frac{  2E_0 W }{mv_F^2 \tau_2}
    \:.
    \end{array}
\end{equation}

Based on Eqs.~(\ref{NS_and_ref_b_c}) one can reformulate the boundary problem for the hydrodynamic flow for the entire sample $0<y<W$.  The effective flow velocity $V_{h} ^{e}(y)$ can be introduced which satisfy the Navier-Stockes equation and the boundary conditions at the very edges $y=0,W$, and is equal to the above result for $V_h(y)$~(\ref{Velocity_hydro}) at $\lambda<y<W-\lambda$. By interpolating equations~(\ref{NS_and_ref_b_c}) to the near-edge regions $0 < y < \lambda$ and $W -\lambda < y < W$ and having in mind~(\ref{VV'}) we obtain in the main order by~$l_2/W$ the boundary problem (\ref{NS_and_ref_b_c_int}).
 In it, the value $l_{sc} = \xi  - \lambda $ is the actual slipping length according to its usual definition via conditions~(\ref{NS_and_ref_b_c_int}) at the very sample edges (see Ref.~\cite{Kiselev2019,Raichev2022}). Since the refined boundary condition describes the difference in the dynamics  of the electrons reflected from the edges and incident towards the edges, the velocity $V_h(y)$~(\ref{Velocity_hydro}) must be greater than velocity~(\ref{V_H_simplest}) obtained from the simplest no-slip conditions $V_h|_{y=0,W} =0$. The latter overestimates the momentum relaxation on the edges, thus the length  $\xi  '$ must be positive. In view of formula~(\ref{xi_re}), this condition is fulfilled at any chosen $\lambda <l_2$.

The semiballistic velocity profiles $V_{b,l}(y) $ and $  V_{b,r}(y) $ in the near-edge layers $0<y<\lambda$ and $W-\lambda <y <W $ are given  by the equation analogous to Eq.~(\ref{Pi2}):
\begin{equation}
 V_b(y)
 =  \Big(
 \int \limits_{-\frac{ \pi}{2} +\delta }^{ \frac{  \pi}{2} - \delta  }   + \int \limits_{ \frac{  \pi}{2}+ \delta  }^{ \frac{  3\pi}{2} - \delta  }
 \Big)
 \,
 \frac{ d \alpha  }{ 2 \pi }  \,
 f _b (y,\alpha)\,
  \sin \alpha
\,.
\end{equation}
A direct calculation yields  that $ V_{b,l}(y) = V_{b,r}(y) = \mathrm{const} (y) $ in the main order by the small parameter $l_2 /W $, and:
\begin{equation}
	\label{Velocity_ballistic}
	V_{b,l/r}(=\frac{1}{2}\, V_h(\lambda) \Big[\,1-\frac{2\delta + \sin(2\delta )}{\pi}\,\Big]
   \:,
\end{equation}
where $\delta =\delta  _\lambda $ [see Eq.~(\ref{lambda_layers_sharp})].

In this way,  the velocity profile $V(y)$ in the whole sample section, $0<y<W$,  is given by the quasi-Poiseuille flow~(\ref{Velocity_hydro}) at $\lambda < y < W - \lambda$ and by the near-edge flow velocity~(\ref{Velocity_ballistic}) at $0<y<\lambda$ and $W-\lambda < y < W $. As it follows from the qualitative estimations for the model with continuous variations of the flow characteristics, the values of $V(y) $ and $E_H(y)$ calculated within the developed layer model are accurate up to numerical coefficients.

\section{ Flow characteristics in first order by $\omega_c$ and Hall effect within ``sharp-edge'' model}
In this section, we calculate the distribution function in the near-edge regions  in the first order by magnetic field. Such function describes the correction to the Hall electric field from this regions. We compare this corrections with the  bulk Hall viscosity correction to the Hall field presented in Appendix~A.

 In the presence of a magnetic field, first, the Hall electric field appears due to arising of the Lorenz force which should be compensated by the Hall field in the considered long sample.
  Additionally, the velocity profile and the expression for the slipping length change. From the symmetry reasons, the last two effects are quadratic in magnetic field. In what follows, we will study the case of a small magnetic field and consider only the linear in $\omega_c~B$ contributions to the distribution functions:
\begin{equation}
  f_{ b,l } ^1   \, \sim \,
  f_{ b,r }^1  \, \sim \,
  f_h^1 \, \sim \, \omega_c
  \:.
\end{equation}
The Hall field in the hydrodynamic region is described  by the $y$-component of the Navier-Stokes equation~(\ref{NS_eq}) with the coefficients proportional to $\omega_c$. The linear in $\omega_c$ part  $f_h^1 (y)$ of the  hydrodynamic distribution function~(\ref{F_hydro}) is:
\begin{equation}
	f_h^{1}(y,\alpha)=-  m v_F^2  \omega_c  \tau_2^2 \, V_h'(y) \, \cos 2\alpha
  \:.
\end{equation}
The correction $f^1 \sim \omega_c$  to the ballistic distribution function proportional to the magnetic field is to be found from the equation:
\begin{equation}
	v_F \, \cos \alpha  \, \frac{\partial f_{l/r} ^ 1 }{\partial y}
-
 \omega_c  \, \frac{\partial f _{l/r} ^1 }{\partial \alpha}
   = 0
\end{equation}
 with the boundary conditions:
\begin{equation}
	\begin{array}{l}
 \displaystyle	
	\big( f_{b,l}^{1,-} = f_h^{1} \big) \,| _ { y = \lambda , \,  \frac{\pi}{2} < \alpha < \frac{ 3\pi}{2}}
   \,,
   \\\\
   \displaystyle	
		f_{b,l}^{1,+} |_{y = 0 , \,  -  \frac{ \pi}{2} < \alpha < \frac{ \pi}{2}}
   =-\frac{1}{2}\int\limits_{\frac{\pi}{2}}^{\frac{ 3\pi}{2}} f_{b,l}^{1,-}(0,\alpha') \cos \alpha' d\alpha'
   \end{array}
\end{equation}
and
\begin{equation}
	\begin{array}{l}
 \displaystyle	
   \big(\, f_{b,r}^{1,+} =f_h^{1} \, \big)  | _ { y = W-\lambda, \, - \frac{ \pi}{2} <\alpha < \frac{\pi}{2} \,}
   \,,
   \\\\
   \displaystyle	
		f_{b,r}^{1,-} |_{y = 0 , \,  \frac{ \pi}{2}   < \alpha < \frac{ 3\pi}{2}   }
= \frac{1}{2}\int\limits_{- \frac{ \pi}{2}}^{\frac{ \pi}{2}} f_{b,r}^{1,+}(W,\alpha') \cos \alpha' d\alpha'
\:.
	\end{array}
\end{equation}

The solution of this problem in the first two orders by the parameter~$l_2/W$ (and the first order in~$\omega_c $) for the left near-edge semiballistic region takes the form:
 \begin{equation}
  \label{res_f_b_lin}
\begin{array}{c}
\displaystyle
	f_{b,l}^{1,-}(y,\alpha) \, = \,
m \omega_c  V_h(\lambda)  (y-\lambda) +
\\
 \\
 \displaystyle
+ \frac{e E_0 \omega_c  (y-\lambda)^2}{2 v_F \cos^3 \alpha} \, -
 \\
 \\
 \displaystyle
- mv_F^2\, \omega_c  \tau_2 V_h'(\lambda)  \cos 2\alpha    \Big[\,  \tau_2 + \frac{y-\lambda}{v_F\cos\alpha}\, \Big]
\end{array}
\end{equation}
and
\begin{equation}
\label{res_f_b_lin2}
\begin{array}{c}
\displaystyle
	f_{b,l}^{1,+}(y,\alpha)=  -  m \omega_c \lambda V_h (\lambda)  +
\frac{ e E_0 \omega_c  y^2}{2 v_F \cos^3 \alpha}
 -
 \\
 \\
 \displaystyle
- mv_F^2  \frac{ \omega_c \tau_2^2 }{3} V_h'(\lambda)
 \, .
\end{array}
\end{equation}
The functions  $ f_{b,r}^{1,\pm} $ are obtained from $ f_{b,l}^{1\pm} $ by substitutions $y \to W-y $ and  $\alpha \to \pi-\alpha $.

Angular integration of Eqs.~(\ref{res_f_b_lin}) and (\ref{res_f_b_lin2}) over the intervals $| \pi/2 - |\alpha| | > \delta (\lambda) $, according to Eq.~(\ref{0}), provide the zero angular harmonic controlling the distribution of the electric charge. The exact form of Eq.~(\ref{0}) with excluding these vicinities of the angles $ \varphi =\pm \pi/2  $ takes the form:
 \begin{equation}
 \label{00}
  f_{m=0} (y)=
  \Big[
\int \limits_{-\frac{ \pi}{2} +\delta }^{ \frac{  \pi}{2} - \delta  }
 +
  \int \limits_{ \frac{  \pi}{2}+ \delta  }^{ \frac{  3\pi}{2} - \delta  }
\Big]
\,   \frac{d \alpha  }{2\pi}  \: f_b (y,\alpha )
\:.
\end{equation}
 This yields the potential $\phi (y) \approx f^{m=0}_{b}(y) $ of the Hall electric field $E_{H} ^b = \phi' (y)  =\mathrm{const}(y) $  in the near-edge  semiballistic regions. In both the left and the right semiballistic layers we have:
\begin{gather}
\label{EHb}
	E_{H} ^b
 =
  \frac{ m \omega_c }{e}
  \int  \limits  _{\frac{\pi}{2}+ \delta } ^{\frac{3\pi}{2} - \delta }
  \frac{d\alpha}{2\pi}
 \Big[    \,  v_F \tau_2 V_h'(\lambda)  \frac{\cos2\alpha} {\cos\alpha} - V_h(\lambda) \, \Big]
 \:.
\end{gather}
In Fig.~\ref{Fig2} we draw the calculated Hall electric field in the near-edge and the central regions.

From Eqs.~(\ref{VV'}) and (\ref{EHb}) it follows that the near-edge contribution  $ \Delta  U_{H,b} = 2 E_{H} ^b  \lambda $ to the Hall voltage $U_H = \phi(W) - \phi(0) $  takes the form:
\begin{equation}
\begin{array}{c}
 \displaystyle
    \Delta  U_{H} ^b = \frac{   2\omega_c  E_0   W \lambda  }{ \tau_2 }\,\Big\{
  \, \frac{\xi}{v_F^2} \, \Big[ \, 1- \frac{  2 \arcsin(\lambda ) }{\pi } \,  \Big]
 \, +
 \\
 \\
 \displaystyle
 + \,
  \frac{\tau_2}{ \pi v_F}
 \Big[ \,
 4 r_\lambda
  +
  \ln
  \Big( \:
  \frac{ \displaystyle  1 -r_\lambda  }
  {\displaystyle  1 +r_\lambda }
  \: \Big)
 \, \Big]
 \Big\}
 \:,
\end{array}
\end{equation}
where $r_\lambda  =  \sqrt{1-\lambda^2 /l_2^2}$. The contribution to $U_H$ from the hydrodynamic region can be presented in the form:
\begin{gather}
\begin{array}{c}
\displaystyle
	U_{H} ^h  \, = \,\frac{ m \omega_c  }{e^2}\, \frac{ I_h  }{n_0} \, + \,    \Delta U_{H} ^h  \: ,
\\
\\
\displaystyle
\Delta U_{H} ^h    = \frac{m}{e^2 }\eta_{xy} [\, V_h' (W-\lambda) -V_h' (\lambda) \, ] =
\\
\\
\displaystyle
=
 -  2 \omega_c \tau_2 W  E_0 \:.
\end{array}
\end{gather}
Now we can express the total Hall voltage $U_H $ in the form containing the total current~$I$, the above  hydrodynamic correction~$\Delta U_{H,h} $, and the ballistic contributions~$ U_{H,b}$:
\begin{equation}
   U_H  \, = \,  \frac{m}{e^2} \,   \omega_c \, \frac{  I }{n_0 }   \,  +   \, U_{H} ^b  \,  + \,   \Delta U_{H} ^h  \,
 - \,    \frac{m}{e^2} \frac{ \omega_c  }{ n_0 }  \,  \Delta  I_b \:,
\end{equation}
where  the last term, $ \omega_c \Delta  I_b /n_0$, arises due to the difference of $I$ and $I_h$ within the current model. The difference $I-I_h$ is the near-edge semiballistic contribution $\Delta  I_b $ determined by Eq.~~(\ref{I}).

\section{Derivation of the differential equations and boundary conditions within ``three-harmonic'' model}
In this section within the method of taking into account three angular harmonics of the distribution function, we present the derivation and the full form of the differential equation for the mean velocity $V(y) $ and momentum flux components $\Pi_{xy,yy} (y)$ as well as for the boundary conditions at the edges $y=0,W$. In the main text, in Section IV, the obtained equations are solved in the limit of a weak magnetic field, $\omega_ c \tau_{2,3} \ll1 $.

The kinetic equation~(\ref{kin3}) in the first two orders by the parameters $ l_2 /W \ll 1  $ and $ l_3 /W \ll 1  $  turns into the ordinary differential equation for the amplitudes of the harmonics~$ f_0 $, $ f_{ms} $, and $ f_{mc} $:
\begin{equation}
 \label{kinetic_3}
\left\{
	\begin{array}{l}
    \displaystyle
    v_F f_{2s}' / 2 \,  - \,  eE_{0} \, = \,  0
  \\\\
 v_F  f_{2c}'  /2  \, + \, v_F  f_{0}' \, - \,  \omega_{c} f_{1s}   \, = \,   0
  \\\\
 v_F  f_{3c}' /2 \,- \, 2 \omega_{c} \,  f_{2s}  \, = \, - \,  f_{2c} /\tau_{2}
  \\\\
 v_F  f_{3s}'/2  \, + \, v_F   f_{1s}' /2 \,+\,  2\omega_{c}  \, f_{2c}  \, = \,  - \,  f_{2s} /\tau_{2}
 \\\\
 v_F  f_{2c}' /2   \, - \,  3\omega_c \, f_{3s}   \, = \,  - \,     f_{3c} /\tau_{3}
 \\\\
  v_F  f_{2s}' /2  \, + \,  3 \omega_c \, f_{3c}
    \, = \,  - \,     f_{3s} /\tau_{3}
\end{array}
\right.
 .
\end{equation}
The second of these equations provides the value of the Hall field  $E_H (y) \approx - f_0' (y) $. Using  the last two of Eqs.~(\ref{kinetic_3}), we obtain the expressions that  allows to get rid of $f_{3c}$ and $f_{3s}$ in other  equations. In the linear order by $\omega_c $ we have:
 \begin{equation}
\label{3}
\begin{array}{l}
 \displaystyle
f_{3c} = - ( l_3  / 2 )  (\, f_{2c}' \, + \, \beta_3\,  f_{2s} ' \, )
 \:,
 \\\\
f_{3s} = - ( l_3  / 2  ) (\, f_{2s}'  \, - \, \beta_3\, f_{2c}' \, )
\:,
 \end{array}
\end{equation}
where  $ \beta_{3} = 3\omega_{c}\tau_{2}  \ll 1   $ and $l_3 = v_F \tau_3 $.

Now we rewrite Eqs.~(\ref{kinetic_3}) and (\ref{3}) in the form containing the usual hydrodynamic variables:
\begin{equation}
V =\frac{ f_{s1}}{mv_F} \, , \quad  \Pi_{yy} = \frac{  f_{c2}}{2} \, ,\quad  \Pi_{xy}= \frac{f_{s2}}{2} \, ,
\end{equation}
in order to get a more transparent form of the Navier-Stokes-like equations. As a result, we obtain:
  \begin{equation}
  \label{total_eq_4_functions}
  \left\{
  	\begin{array}{l}
 \displaystyle
 e E_{H}  = - m \omega_{c} \, V +  \Pi_{yy}'
 \\\\
 \displaystyle
\Pi_{xy}'  = e E_{0}
\\\\
 \displaystyle
\Pi_{xy}= - m \eta  _0  \, V' - \beta_2\,  \Pi_{yy} - \lambda^2  \beta_3 \, \Pi_{yy}''
\\\\
\displaystyle
\Pi_{yy} =  \beta _{2} \, \Pi_{xy} + \lambda^2  \,\Pi_{yy}''
\end{array}
\right.
.
\end{equation}
Here  we introduced the length parameter
\begin{equation}
 \lambda = \frac{\sqrt{l_2l_3 }}{2 }\, ,
\end{equation}
the viscosity in zero magnetic field $ \eta  _0 = v_F  ^2 \tau_{2} / 4 $, the parameter $   \beta _2 = 2\omega_{c}\tau_{2}  \ll 1  $  and used that $\Pi_{xy}'' =0$ [see the second of equations~(\ref{total_eq_4_functions})]. These equations govern the Poiseuille flow in the three-harmonic approximation.

Equation~(\ref{total_eq_4_functions}) can be rewritten as a close system of two differential equations of the second order for the functions $V(y)$ and $\Pi_{yy}(y)$:
 \begin{equation}
 \label{two}
  \left\{
  	\begin{array}{l}
 \displaystyle
  m \eta_0  V'' = - (  \beta_2\,  +\beta_3 )\, \Pi_{yy}'
-  (1 - \beta _{2}  \beta _{3} )  eE_0
\\\\
\displaystyle
\lambda^2  (1 - \beta _{2}  \beta _{3} )  \Pi_{yy}'' =
  (1+\beta_2^2 )  \Pi_{yy}   +  m \eta  _0  \beta_2   V'
\end{array}
\right.
.
\end{equation}
Thus  the values $E_H(y)$ and $\Pi_{xy}(y)$ in Eqs.~(\ref{total_eq_4_functions}) plays the role of auxiliary variables which can be excluded by the second and the  third of equations~(\ref{total_eq_4_functions}).

Now let us formulate one of proper boundary conditions for the final hydrodynamic-like equations~(\ref{two}) for $V(y)$ and $\Pi_{yy}(y)$.  Based on the mathematical structure of these equations, they are to be supplemented by the boundary conditions at the edges $y=0,W$ for some two linear combinations of the functions $V$ and $\Pi_{yy}$ and their derivatives at the two edges.

According to the physical essence of function~(\ref{F_hydro_3}) and equations~(\ref{two}), the exact diffusive conditions~(\ref{diff_b_c}) and  (\ref{diff_b_c2}) for full distribution function cannot be satisfied. So there is a variety of possible approximate boundary conditions for the truncated distribution~(\ref{F_hydro_3}), those are successors of the exact ones~(\ref{diff_b_c}),~(\ref{diff_b_c2}). We will seek for the boundary conditions for $V,\, V', \, \Pi_{yy} , \, \Pi_{yy} ' $ at $y=0,W$ considering that the values
$
  \Pi_{xy} , \, f_{3c} , \, f_{3s}
 $
   at $y=0$ and $y=W$  are related to
  $
    V,\, V', \, \Pi_{yy} , \, \Pi_{yy} '
   $
     by equations~(\ref{3}) and the  third of equations~(\ref{total_eq_4_functions}).

To construct such approximate  condition, one can use different approaches. One of them is as follows. At each edge $y=0,\,W$, for a given set of $ f_{3c} $, $ f_{3s} $ and $\Pi_{xy}$, we find those $ V$ and $\Pi_{yy} $, for which boundary conditions~(\ref{diff_b_c}) and  (\ref{diff_b_c2})  are most closely satisfied in sense of minimization of some discrepancy function related to Eqs.~(\ref{diff_b_c}) and~(\ref{diff_b_c2}). Then we use interconnection of these values via  equations equations~(\ref{3}) and (\ref{total_eq_4_functions}). As a measure of the discrepancy, we consider the following quadratic functions $ \Delta_l = \Delta_l (V, \Pi_{yy} )$  at the left edge:
\begin{equation}
\label{residue}
\Delta_l = \int\limits_{-\pi/2}^{\pi/2}  \Big[f(\alpha)+\dfrac{1}{2}\int\limits_{\pi/2}^{3\pi/2} f(\alpha') \cos \alpha' d\alpha'\Big]^2 d\alpha \: ,
\end{equation}
and similar function $ \Delta_r = \Delta_r (V, \Pi_{yy} )$   at the right edge:
\begin{equation}
\label{residue2}
\Delta_r = \int\limits_{\pi/2}^{3\pi/2}\Big[f(\alpha)-\dfrac{1}{2}\int\limits_{-\pi/2}^{\pi/2} f(\alpha') \cos \alpha' d\alpha'\Big]^2 d\alpha \: .
\end{equation}

The values of $V $ and $  \Pi_{yy}  $ at the boundaries, which correspond to the truncated distribution function (\ref{F_hydro_3}) with given $f_{3s}$,  $ f_{3s} $ and $\Pi_{xy}$, are those for which functional $\Delta_l$ and $\Delta_r$ take its minimum value.  The corresponding  Euler equations for this variational problem  for the left edge function~(\ref{residue}) are linear and  take the form:
\begin{equation}
\label{conditions_3_start}
  \dfrac{\partial \Delta_l (V ,  \Pi_{yy}  ) }{\partial V } = 0
  \,,\quad
 \dfrac{\partial \Delta_l (V ,  \Pi_{yy}  ) }{\partial  \Pi_{yy} }  = 0
\:.
\end{equation}

Calculations based on Eqs.~(\ref{F_hydro_3}) and~(\ref{residue}) yield:
\begin{equation}
\label{conditions_3_res}
\displaystyle
 \Big( m v_F  V  + \frac{16}{3\pi} \,  \Pi _{xy  } \Big)
 \Big|_{y=0} =0
    \, ,
 \end{equation}
\begin{equation}
\displaystyle
 \Big(      \Pi_{yy} + \frac{64 }{55 \pi    }   f_{3c} \Big)
 \Big|_{y=0} =0
  \: .
\end{equation}
For the right edge $y=W$ the similar boundary condition follows from minimization of function~$\Delta _r(V,\Pi_{yy})$~(\ref{residue2}):
\begin{equation}
\displaystyle
  \Big( mv_FV  - \frac{16}{3\pi} \,  \Pi _{xy  } \Big)
  \Big|_{y=W} =0
    \, ,
\end{equation}
\begin{equation}
    \label{conditions_3_res2}
\displaystyle
    \Big(      \Pi_{yy} -  \frac{64 }{55 \pi    }  f_{3c} \Big)
    \Big|_{y=W} =0
   \:.
\end{equation}
The values $\Pi_{xy}$ and $f_{3c}$ in above equations (\ref{conditions_3_res})-(\ref{conditions_3_res2}) should be expressed via $V$ and $\Pi_{yy}$ by the first of equations (\ref{3}) and the third and second equations~(\ref{total_eq_4_functions}).

The resulting four boundary conditions (\ref{conditions_3_res})-(\ref{conditions_3_res2}), containing $V$, $V'$, $\Pi_{yy}$, and $\Pi_{yy}'$ a sufficient to find a proper solution of
the system (\ref{two}) of second-order differential ordinary equations.
   In the main text we construct such solution for the limit of weak magnetic fields.

\bibliography{b}

\end{document}